# Electrohydrodynamic instability of Cu, W and Ti metal nanomelts under radiofrequency E-fields from multiphysics molecular dynamics simulations with coarse-grained density field analysis

Shangyong Wu[a#], Rui Chu[a#], Wenqian Kong[a], Hongyu Zhang[a], Le Shi[a], Kai Wu[a], Yonghong Cheng[a], Guodong Meng[a*], Bing Xiao[a*]

a: School of Electrical Engineering, State Key Laboratory of Electrical Insulation and Power Equipment, Xi'an Jiaotong University, Shaanxi, 710049, P.R. China
# Shangyong Wu and Rui Chu contributed equally to this manuscript
* Corresponding Authors: gdmengxjtu@xjtu.edu.cn (Guodong Meng); bingxiao84@xjtu.edu.cn (Bing Xiao)

**Abstract**: Employing both electrodynamics coupled with molecular dynamics (ED-MD) simulations for atomistic models and the dynamic instability theory of electrocapillary wave, we investigate the structure evolutions and thermal runaway process of Cu, Ti and W nanotips with radii of curvature of 1 nm and 5 nm under various radiofrequency electric field conditions. The associated critical parameters including the critical electric field, spatial and temporal scales of the electrohydrodynamic instability of molten apexes are obtained by proposing the workflows that utilize the atomistic models in ED-MD simulations to calculate kinematic viscosity tensor components and mass density spatial distributions for the nanomelts with electric fields. Our current ED-MD simulations for nanotips show a non-monotonical variation of the time delay versus the electric field frequency for metal nanotips, and the presence of a critical rf electric field amplitude triggering the thermal runaway regardless of the field frequency. The calculated mass densities and kinematic viscosities of nanomelts for metal nanotips are found to be drastically different to those of bulk liquid metals at the melting point. Specifically, the viscosity of nanomelt under the rf electric field is revealed to be several orders of magnitude higher than the bulk liquid metal, resulting in substantial increase of spatial and temporal scales in the instability theory of electrocapillary wave within the viscosity-dominated regime, compared to the results of ED-MD simulations for Cu and Ti metals, while good agreement between the two methods on the critical wavelength and time delay of thermal runway is found for W nanotips.

# 1 Introduction

The structural instability of metal micro- and nano-protrusions under the applied strong electric field is critical to understand the underlying physics for the initiation of electric breakdown for many high-voltage vacuum electrical and power equipment as well as the vacuum nano-electronic devices such as field emitters,[1–3] vacuum field emission transistors (vacFET)[4–6] and field emission diodes.[7–9] It is now well recognized that metal protrusions could also contribute significantly to the electric breakdown process of large particle accelerators, greatly threatening the operational performances of those high vacuum particle colliders, i.e., the large compact linear collider (CLIC) at CERN.[10–13]

Investigating the structural instability of metal protrusions dynamically or in-situ within the current state-of-the-art experimental tools such as scanning electron microscope (SEM) and transmission electron microscope (TEM) under the high electric field and strong heating process remains very difficult especially for the micro- or nano-size sharp metal tips because the limitations in the spatial and temporal resolutions of those tools, and the strong electromagnetic interference with the instrument during the electric breakdown process. Nevertheless, the morphologies, phase transitions and microstructures of metal nanotips or nano-protrusions during the field emission process or before and after electrical breakdown can be characterized experimentally, and which have been done in many previous studies.[14–20] Meanwhile, computer simulations based on either electrohydrodynamic simulation or electrodynamics coupled with molecular dynamics (ED-MD) simulation could overcome most of those issues, providing critical structure evolution details and probably underlying mechanisms regarding the structural instability of materials in time scale ranging from ns ($10^{-9}$ s) to ps ($10^{-12}$ s) under strong E-field.[11,21–30] Notably, typical multiphysics simulations such as electro-hydrodynamics (EHD) and magnetohydrodynamics (MHD) based on finite element method are routinely employed to elucidate the mass and heat transport properties of macroscopic and micro-size metal protrusions with the applied electric field.[11,23–27] However, both applicability and validity of the finite element based multiphysics simulations to metal protrusions at nanoscale remain an open question, because the thermophysical properties and hydrodynamics of solids and liquids at nanoscale are expected to be different to those of bulk state even without the external electric field.[31–34] Otherwise, some recent advancements in the development of multiscale-multiphysics hybrid molecular dynamics simulations enabled the direct modelling of structural deformations (plastic and elastic), phase transitions among solid, liquid and vapor states, and thermal evaporation (thermal runaway event) in atomistic models for metal nanotips and nano-protrusions that are dynamically coupled with local electric fields both spatially and temporally.[28,29,35–37] ED-MD simulation provides a promising complementary methodology to those conventional EHD and MHD developments in addressing the electrical breakdown phenomena for various materials at sub-micrometer and nanoscale under locally high electrical fields and strong heating process. Obviously, the combination of EHD/MHD with hybrid molecular dynamics simulation such as ED-MD has the great potential to merge the spatiotemporal gap between those methods, probably providing a unified methodology and workflow to reveal the mass and heat transfer processes with electrical field for metal nanotips or nano-protrusions in the future.

Based on previous experimental studies and multiscale-multiphysics simulations (EHD/MHD/ED-MD approaches) for micro- and nano-protrusions, the structural instability of sharp metal tips under the strong local electrical fields is very often triggered by the thermal runaway event at the apex region, especially when the apex is in the molten state due to the intense field emission process and the high resistive heating mechanism.[23–27] However, the structural evolution of micro-protrusions under the high E-field from EHD or MHD multiphysics simulations seems to be different to that of nanotips and nano-protrusions using ED-MD hybrid molecular dynamics simulations. For micro-protrusions, the EHD and MHD simulations typically illustrate the gradual formation of conic cusps known as Taylor cone at the apex region for the molten liquid metals,[23,27,38,39] further enhancing the local electrical field distortions and field emission process. Eventually, the Taylor cone like sharp protrusion of molten metal fully collapses either through the explosive evaporation mechanism or by emitting the charged droplet jet.[27,38] The instability of those Taylor cones, which is referred to Tonks-Frenkel (TF) or Lamor-Tonks-Frenkel (LTF) instability, is related to the dedicated balance among three forces including the electrostatic forces in terms of Maxwell stress, capillary and gravitational forces.[40,41] On the other hand, the gravitational force is not expected to play significant role in structural evolution of nanotips or nano-protrusions with the strong local electric fields in molecular dynamics simulations, because the gravitational potential is negligible in comparison to surface tension for molten metal at atomic scale. Previously, ED-MD simulations have been conducted for nanotips and nano-protrusions consisting of various metals such as Ti,[36] Cr,[42] Cu,[28,37,42] Mo[35] and W[35] to study their atomic structural evolutions and the thermal runaway behaviors under the strong E-field and intense field electron emission. In most of those investigations, the structural evolution of metal nanotips or nano-protrusions in the thermal runaway even undergoes three stages including the melting of apex region with strong Joule heating process (due to field emission), the large mechanical deformation of the molten region in terms of stretching and necking, and the final explosive thermal evaporation process.[36] Interestingly, the thermal evaporation of metal nanotips, as predicted from ED-MD hybrid molecular dynamics simulations, is found to proceed by detaching portion of stretched molten metal due to electrostatic forces under strong E-field for metals with low melting points (Cu, Zn, Au and etc.), probably resembling that of charged droplet jet in TF or LTF instability for micro-protrusions or bulk liquid metals. In fact, Zhang and coworkers have studied the atomic structure evolutions of Cu and Cr nanotips with a radius of curvature of 3 nm using their in-house ED-MD method, directly observing the formation of cusp-type sharpening structure for both metal nanotips.[42] However, Ref. 42 assumes an axial symmetry for the electrostatic forces acting on the surface atoms of metal nanotips, leading to the open question that whether the predicted Taylor cone like apex could persist after lifting the applied constraints for electric forces in axial direction. Meanwhile, the thermal runaway of nanotips containing refractory metals such as W was mainly attributed to the formation of sharp atomistic protrusions at the apex of the nano-protrusion and the subsequent field evaporation of few atoms or atomic clusters with high local electric fields based on our previous ED-MD simulations.[35] Such thermal runaway process differs substantially to the TF or LTF instability of metal micro-protrusions with the applied E-field. So far, no work has been done to establish the direct connection between instability of atomistic nano-protrusions under the E-field from ED-MD simulations and the Taylor cone instability of TF or LTF type in electro-hydrodynamics.

Our current work is partly motivated by a series of seminar works conducted by Zubarev and coworkers using either the electrohydrodynamic simulations or the wave mechanics to study the growth of conic cups on the conductive liquid under the applied E-field.[40,43–46] Those works have discovered two different types of Taylor cone instabilities for metal liquid with the applied electric field, i.e., the conventional conducting fluid at macroscopic scale and the viscous conducting liquid at nanoscale. For the former conducting liquid, the limiting opening angle for the Taylor cone was known to be 98.6°, while the metal liquid of nano-protrusions gave a significantly smaller limiting angle of 33.1° when kinematic viscosity of molten region plays the decisive role in determining the instability of electrocapillary waves on the surface of conducting liquid. In addition, the results implied that electric field induced structural instability of metal nanotips or nano-protrusions could develop the unusual sharpened conic cusps at the apex with the greatly reduced conic opening angles, mimicking the atomistic protrusions formed on the apex of nanotips or nano-protrusions under high electric field in ED-MD simulations. Otherwise, Ref. 40 also derived the characteristic spatial and temporal scales for the dominate instability mode of electrocapillary waves for conducting liquid in two different regimes, i.e., the ideal conducting liquid and the viscosity-dominated regimes, respectively. Moreover, the boundary between those two regimes in the electro-hydrodynamics of the molten nano-protrusions were also defined by the relationships that are the explicit functions of some fundamental thermophysical parameters of conducting liquid such as surface tension ($\alpha$), kinematic viscosity ($v$) and density ($\rho$). All those parameters could be also calculated from the atomic trajectory of typical multiscale-multiphysics molecular dynamics simulations such as ED-MD method. Therefore, it is now possible to establish direct correlation between atomistic model and the electro-hydrodynamics of conducting liquid for the characteristic spatial and temporal scales in the TF or LTF instability of nanotips or nano-protrusions under the electric field.

In this paper, we have conducted a series of ED-MD simulations for nanotips of three transition metals including Cu, W and Ti under the radiofrequency electric field using the in-house multiscale-multiphysics hybrid molecular dynamics simulation program known as FEcMD code.[47] Employing the coarse-grained mass density analysis, the atomic structure evolutions and the mass density distributions of those metal nanotips with different radii of curvature were obtained at various stages during the thermal runaway event under the critical E-field amplitude and frequency. From the calculated mean density of molten apex and the kinematic viscosity of each metal nanotip, some key parameters for characterizing the TF instability in the viscosity dominated regime were evaluated for the nanotips of three metals for the first time using ED-MD simulations. Our current work provided an important step to bridge the existing methodology gap between the continuous electro-hydrodynamics and multiphysics discrete particle dynamics regarding the TF instability of nanotips or nano-protrusions with the applied electric field.

## 2 Computational methods and details

**2.1 ED-MD simulations**

All ED-MD simulations were carried out using the in-house multiscale-multiphysics program known as FEcMD code.[47] Details regarding methodologies and workflow of FEcMD code can be found in Ref. 47. We performed ED-MD simulations for conical metal nanotips of Cu, W and Ti elements with two different radii of curvature, i.e., $r_0 =$ 1 nm and 5 nm, respectively. The three metals were particularly chosen in this work because they exhibit different crystal structures and thermophysical properties. Specifically, Cu (W) form cubic crystal structure in terms of face-centered cubic or FCC (body-centered cubic or BCC) lattice, while Ti element has the hexagonal close-packed (hcp) lattice structure at ambient conditions. Besides the crystal structures, other important thermophysical properties of Cu, W and Ti are summarized in Table. 1, and which were employed in ED-MD simulations for the relevant element. The initial configuration of the nano-protrusion in ED-MD simulation is illustrated in Figure 1. Each nano-tip consists of two segments, i.e., the upper half atomistic model (50 nm) and the lower half coarse-grained model (50 nm). The entire nanotip has a total height of 100 nm, and which was fixed onto the coarse-grained substrate representing the bulk region of metal. The finite element meshes were generated for the whole simulated space with specialized algorithms to differentiate the vacuum, surface and bulk regions of a nano-protrusion. For metal nanotip, the electrodynamics simulation was conducted for vacuum and surface mesh points, while the MD simulation only applied to the atomistic model. This multiscale modelling strategy was employed to mainly reduce computational costs for molecular dynamics simulations. The two-temperature model (TTM), which involves the heat exchanging mechanism between electron and phonon subsystems, was used to simulate the heat conduction in the nano-protrusion. Employing TTM is necessary here to more accurately describe the dual-channel heat conduction processes in materials under radiofrequency electric field, because the heat energy could dissipate through the non-thermal mechanism such as high-frequency electronic heat conduction process.[48,49]

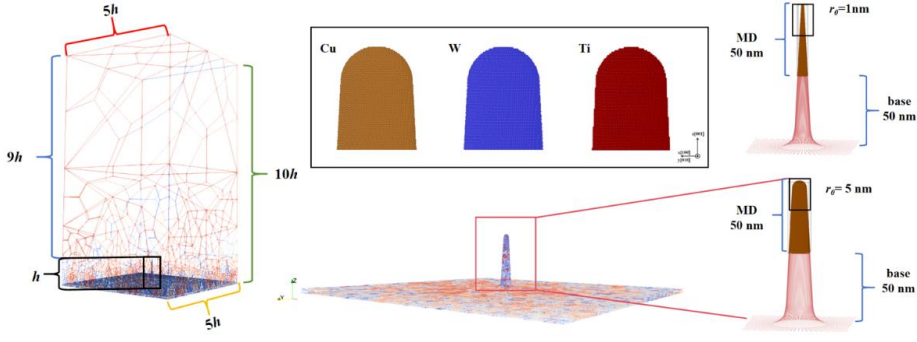

**Figure. 1** The atomistic and continuum domains in the multiscale ED-MD simulation of metal nanotip.

Table 1. Crystal structures, initial nanotip parameters and other thermophysical properties of Cu, Ti and W nano-protrusions employed in ED-MD simulations. Here, $\Phi$, $C_{ep}$ and $C_V$ are the work function, electron-phonon energy exchange rate and the isothermal specific heat, respectively. $H_0$ refers to the initial height of the atomistic model and $r_0$ denotes the initial radius of the curvature for each nanotip.

| Metal | Lattice type | Nanotip parameters | $\Phi$[50] (eV) | $C_{ep}$[51,52,53] W/(K·m³) | $C_V$[52] J/(cm³·K) |
|---|---|---|---|---|---|
| Cu | FCC | $H_0$ = 50 nm, $r_0$ = 1 nm | 4.59 | $1 \times 10^{17}$ | 3.45 |
|  |  | $H_0$ = 50 nm, $r_0$ = 5 nm |  |  |  |
| Ti | HCP | $H_0$ = 50 nm, $r_0$ = 1 nm | 4.33 | $1.3 \times 10^{18}$ | 2.36 |
|  |  | $H_0$ = 50 nm, $r_0$ = 5 nm |  |  |  |
| W | BCC | $H_0$ = 50 nm, $r_0$ = 1 nm | 4.63 | $2 \times 10^{17}$ | 2.59 |
|  |  | $H_0$ = 50 nm, $r_0$ = 5 nm |  |  |  |

For MD simulations, the maximum duration was set to 200 ps with the time step of 1 fs. The ED calculations were conducted using a time step of 0.5 fs to update the electric field values on mesh points. Meanwhile, we chose a time step of 20 fs to calculate the electron and phonon temperature profiles in the interior of nano-protrusion. The interatomic interactions of Cu, Ti and W nano-protrusions were described by the corresponding EAM potentials, i.e., Sabochick-Lam-Mishin EAM (Cu),[54] Olsson EAM (W)[55] and Ackland EAM (Ti).[56] The initial temperature of all nano-protrusions was set to 300 K for electron and phonon subsystems. Frequencies were given as 1 GHz, 10 GHz, 20 GHz, 40 GHz, 60 GHz ,80 GHz and 100 GHz for the radiofrequency E-fields in ED-MD simulations. Otherwise, the amplitude of radiofrequency electric fields was ranged from 200 MV/m to 2 GV/m. In the current implementation of radiofrequency electric fields in FEcMD code, the truncated sine waveform in the positive half period was adopted with user specified electric field amplitude and frequency.

**2.2 Coarse-grained mass density field analysis**

In order to calculate some critical parameters such as critical E-field strength, critical wavenumber and increment of instability that characterize the TF instability for metal nanotips from the thermophysical properties including the mass density of molten apex, kinematic viscosity and surface tension, we have developed the coarse-grained mass density field analysis to extract the instantaneous mass density of the metal liquid at the apex of nanotips using the particle trajectory of ED-MD simulations. In the coarse-grained scheme, the point-like mass of each particle is smeared in the whole space through a spatially normalized distribution function, i.e., the Gaussian distribution function as given by Eq. (1), following the method proposed by Willard and Chandler in Ref. 57. Here, the spatial position is referred to $r$, and $d$ denotes the dimension of the system; $\xi$ represents the coarse-grained length of the point-like mass. Although, the choice of $\xi$ is expected to be dependent on the system under consideration, the value was set to 2.4 Å in this work, the same as that of Ref. 57. Based on the spatially normalized Gaussian distribution of particle mass, the coarse-grained density field was calculated from Eq. (2) for each instantaneous snapshot in the ED-MD simulation, where the summation goes over all particles (*i*) in the atomistic model.

$$\phi(\boldsymbol{r};\xi) = (2\pi\xi^2)^{-\frac{d}{2}} \exp(\frac{-r^2}{2\xi^2}) \qquad (1)$$

$$\bar{\rho}(\boldsymbol{r},t) = \sum_i \phi(|\boldsymbol{r}-\boldsymbol{r}_i(t)|;\xi) \qquad (2)$$

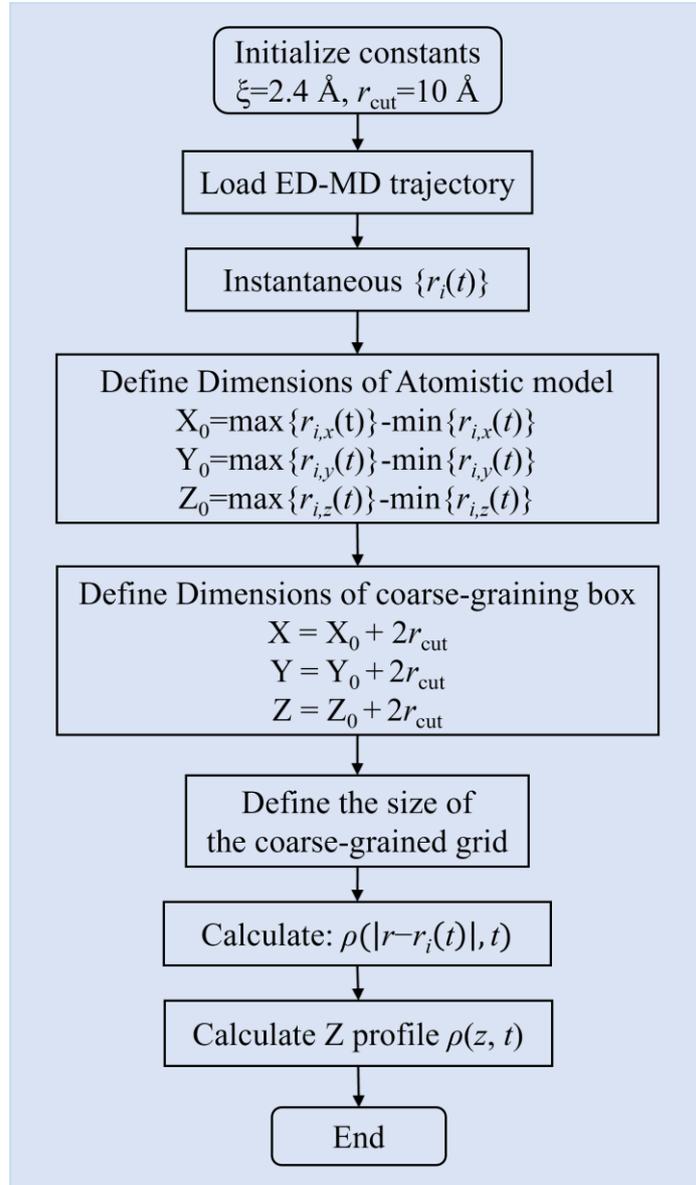

**Figure. 2** The workflow of coarse-grained mass density field analysis.

The workflow of the coarse-grained density field that is employed in the current work is illustrated in Figure. 2. The implementation of the method typically involves the steps as follows: extracting the instantaneous atomic coordinates from the trajectory of ED-MD simulation and determining the dimensions of the coarse-grained domain that covers the whole atomistic model of nanotip; the coarse-grained volume is discretized by a rectangular mesh grid of the size $N_x \times N_y \times N_z$, where the values of $N$ are chosen according to the lengths of coarse-grained domain in *x*, *y* and *z* directions, especially the large number of grid points may be needed in the *z*-direction (or the axial direction) for a better spatial resolution of mass density field, as shown in Figure. 3(a); the primary grid points are further screened based on their distances with the surface atoms in the atomistic models, removing those abundant gird points that are positioned outside the critical cutoff distance in the coarse-grained domain, and this step is aimed at improving the computation efficiency (See Figure. 3(b)); finally, Eqs. (1)-(2) are used to calculate the mass density at each grid point in the coarse-grained domain.

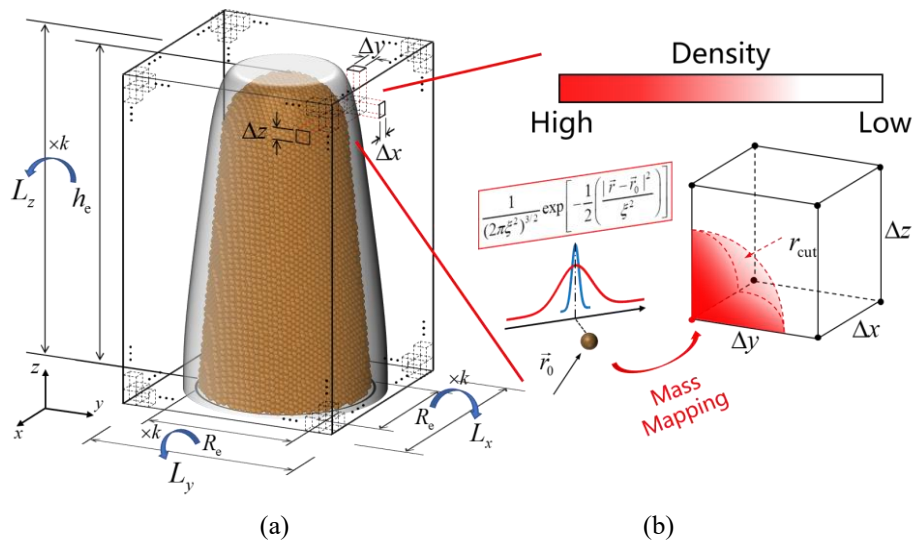

**Figure. 3** Schematic illustrations showing the two main steps in the continuum mass density field analysis: (a): discretization of volume in the atomistic domain with a rectangular mesh grid; (b): the calculation of local mass density field for each grid point.

In our current calculations, the grid spacings were given as $\Delta x = \Delta y = \Delta z = 1.0$ Å, while the cutoff distance in calculating mass densities from surrounding atoms for each grid point was set to 10.0 Å. To evaluate the mean mass densities in the z-direction (or the axial direction of nanotip), Eq. (3) was used. Here, $N_{xy}$ denote the total number of grid points located on the same z-plane in the axial direction, as shown in Figure. 3(a). The resultant is known as the mass density z-profile, and which gives the mean density of the metal nanotip in the atomistic model for different phases, i.e., the crystalline solid, the molten phase and amorphous phase. Obviously, the coarse-grained mass density could be directly related to the continuum mass density field of conventional hydrodynamic simulation. Meanwhile, the radially averaged mass density in the axial direction is considered as a useful mean quantity to characterize the apparent mass density of nanotip, and which could be used to evaluate the critical spatial and temporal scales that may trigger the instability of electrocapillary wave of nanotip apex in the molten state, collaborating with the calculation of kinematic viscosity of the liquid from ED-MD simulation of atomistic model.[40]

$$\rho(z) = \frac{1}{N_{xy}} \sum_{x,y} \bar{\rho}(x,y,z) \qquad (3)$$

### 2.3 Kinematic viscosity of molten apex

Since ED-MD simulation employs the atomistic model to realize the atomic structure evolution of metal nanotip under the influence of heat and electric field, it is straightforward to evaluate the dynamic viscosity of electrified liquid at the apex of the tip by Einstein-Helfand formula, using Eq. (4). Here, $V$, $T$ and $t$ denote the volume of liquid, absolute temperature (K) and the accumulated time step of MD simulation, respectively; the instantaneous particle momentum and positions are given by $p$ and $r$ for the $i$-th atom in the atomistic model, while the $\mu$ and $v$ refer to directions in Cartesian coordinates.

$$\eta_{\mu v} = \frac{1}{Vk_{\mathrm{B}}T} \lim_{t \to \infty} \frac{1}{2t} \left\langle \left( \sum_{i=1}^{N} p_\mu^i(t) r_v^i(t) - p_\mu^i(0) r_v^i(0) \right)^2 \right\rangle \quad (\mu \neq v) \quad (4)$$

Although the use of Eistein-Helfand formula to calculate the dynamic viscosity is questionable for the periodic boundary conditions, the quasi-periodic configuration of nanotip in the atomistic domain may justify it applicability to the simulation model displayed in Figure. 1. The kinematic viscosity is related to the dynamic viscosity by Eq. (5) through the mass density.

$$v = \frac{\eta}{\rho} \quad (5)$$

Since the volume of the molten region at the apex of nanotip varies constantly with the time in the ED-MD simulation, the reliability of applying Eqs. (4) and (5) to evaluate the dynamic and kinematic viscosities of nanofluid critically depends on the accurate extraction of the instantaneous molten region from atomistic model in the MD trajectory. In our current work, this task is achieved through a Python automation script interfaced to the OVITO software.[58] The overall workflow of this automated process is displayed in Figure. 4. The main steps in the extraction process are as follows: importing the MD trajectory into OVITO software as the non-periodic structures; employing the coordination number analysis in OVITO software to differentiate atoms in molten region and surface from those of bulk region in crystalline form; removing all atoms in the crystalline region, and performing the Voronoi polyhedral analysis to the remaining atoms in the nanotip within the same software; removing the surface atoms due to the Voronoi polyhedral construction; performing the relevant calculations for all remaining atoms in the molten region for a designated snapshot in the MD trajectory.

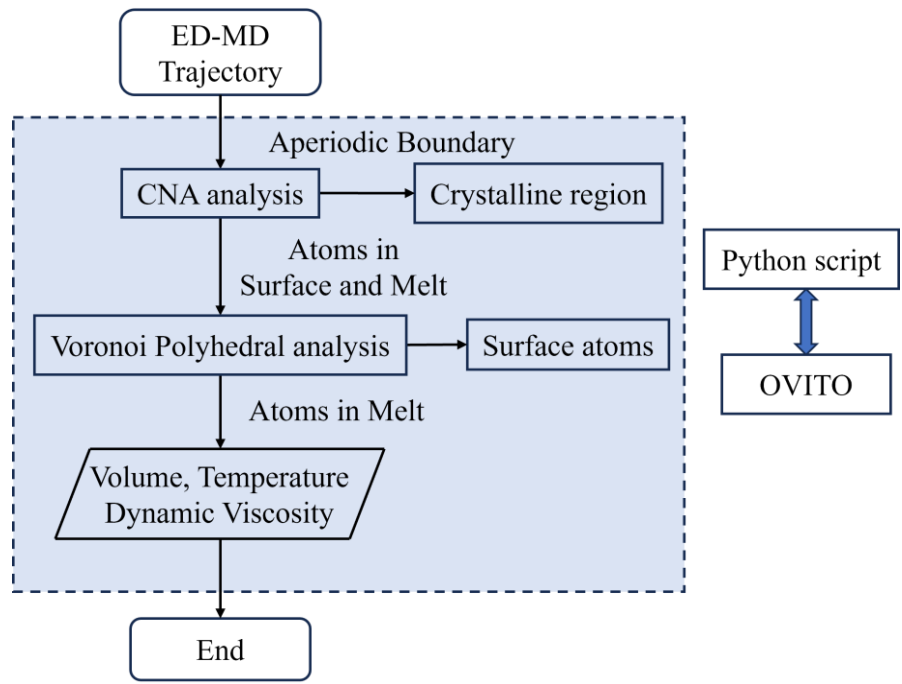

**Figure. 4** Main steps in a Python automation script to extract atoms in the molten region from each snapshot in MD trajectory within OVITO software.

Finally, we would like to emphasize the approach proposed in Figure. 4 for the extraction of the melts from the atomistic domain in ED-MD simulations for metal nanotips is expected to be highly reliable when the apex of nanotip could constantly attain the highly disordered atomic configurations under the intense heating and strong electric field stress before occurring the thermal runaway event. Otherwise, the rapid cooling of the melt at the apex during the negative half waveform may lead to the formation of amorphous form known as glassy state for the metal nanotip, leading to the expected uncertainty or error in extracting the volume of melt from the atomistic model solely based on the local coordination number analysis.

## 3 Results and discussion

### 3.1 Structural evolutions of metal nanotips

Here, we first explore the variations of atomic structures of Cu, Ti and W nanotips with different radii of curvature under various radiofrequency electric fields using ED-MD simulations. Note that the initial geometries and sizes of metal nanotips are summarized in Table. 1 for the three metals. The evolutions of Cu, Ti and W nanotips with the radius of curvature of 1 nm are illustrated in Figure. 5 for the three exemplary electric field frequencies including 10 GHz, 40 GHz and 100 GHz, while the applied external electric field amplitudes are adjusted for each type of metal nanotip to improve the visualization of the characteristic structural deformations and phase transformations at the apex. It is also worth noting that the maximum runtime for each ED-MD simulation is set to 200 ps, assuming no thermal runaway event occurs. However, the ED-MD simulation is terminated automatically by the FEcMD program once the thermal runaway event is detected, i.e., a single atom or atomic cluster is detached from the nanotip.

As shown in Figure. 5(a), the structural deformation of the atomistic model of Cu nanotip under the electric field behaviors drastically different to that of either Ti (Figure. 5(b)) or W (Figure. 5(c)) with the radius of curvature of 1nm. Cu nanotip is substantially blunted through the formation of the large mushroom-head like molten region because of the strong resistive heating that is caused by the intense field electron emission process. Since, the Cu nanotip shows higher total thermal conductivity than those of Ti and W, the resistive Joule heat and Nottingham heat are more quickly dissipated into the bulk region for Cu nanotip, compared to either Ti or W case. Consequently, Cu nanotip with a radius of curvature as 1nm undergoes an obvious recrystallization during the cooling process, i.e., the time interval corresponds to the truncated negative half waveform of the radiofrequency electric field. This is evident from the common neighbor analysis (CNA) displayed in Figure. 5(a), where the red lines indicate the formation of a large number of stacking faults at {111} crystallographic planes for FCC-Cu metal. Stacking faults are also seen in the case of Ti nanotip with the HCP crystal structure within the (0001) plane at the molten apex during the recrystallization process (See Figure. 5(b)). Similar stacking fault is almost absent in the W nanotip, as can be seen from Figure. 5(c). Otherwise, the applied electric field amplitude and frequency also significantly impact the recrystallization of molten apex in the case of Cu nanotip. For example, when the applied electric field amplitude is below 650 MV/m, the molten apex could fully transform into the crystalline solid during the cooling stage at 10 GHz and 40 GHz. However, the apex of Cu nanotip remains partly in the molten state when the frequency of electric field is set to 100 GHz under the amplitude of 650 MV/m. This observation also applies to the Ti nanotip where the apex region is fully crystallized at lower frequency (10 GHz) and electric field amplitude (250 MV/m), compared to those of Cu nanotip. Those results can be understood from the fact that the cooling time interval between two consecutive heating periods is gradually shortened with the increase of electric field frequency, heat dissipation processes through phonons and electrons are eventually overwhelmed by the heating mechanisms (Joule and Nottingham heats), leading to the net increasing of electron and phonon temperatures in ED-MD simulations for all metal nanotips at apex. Besides the melting of apex for all three metal nanotips, the deformation of Cu nanotip under heat and electric field also differs from other two metals. As shown in Figure. 5(a), bending and tilting of Cu nanotip are very often observed in all ED-MD simulations, accompanying many stacking faults in the deformed region. For Cu nanotip, those stacking fault lines at {111} crystallographic planes are not aligned within the horizontal (001) plane, their formation naturally leads to the bending of the tip that is initially oriented in [001] direction. Notably, Ref. 36 studied the Cu nanotips with three different crystallographic orientations in the axial direction such as [001], [110] and [111], and the bending deformation was not seen for nanotip oriented in [111] direction as the stacking fault plane is perfectly aligned in the horizontal direction (radial direction). In fact, bending is not seen in Ti nanotip oriented toward [0001] direction, because the stacking faults are now within the horizontal (0001) plane. Meanwhile, neither bending deformation nor stacking fault is seen in the case of W nanotip with a radius of curvature of 1 nm, probably because the stacking fault formation energy of W even at the most compacted (110) crystallographic plane is significantly higher than that of either Cu or Ti, i.e., 773 mJ/m$^2$ (W, (110) plane), 338 mJ/m$^2$ (Ti, (0001) plane) and 180 mJ/m$^2$ (Cu, (111) plane), respectively.[59,60] Therefore, the formation of stacking faults during the cooling stage is likely impeded in W nanotip.

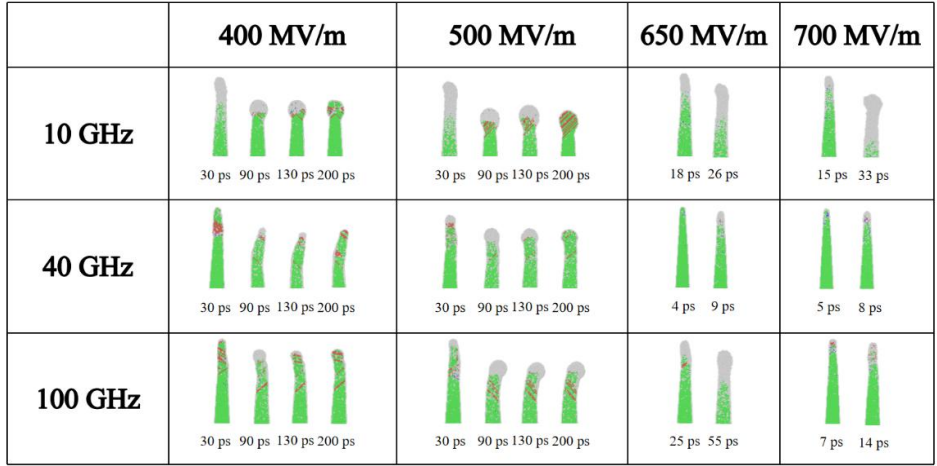

(a)

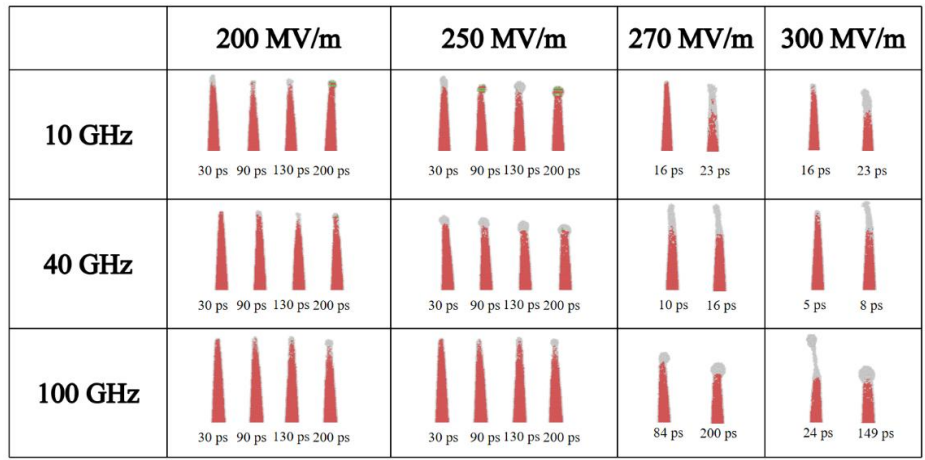

(b)

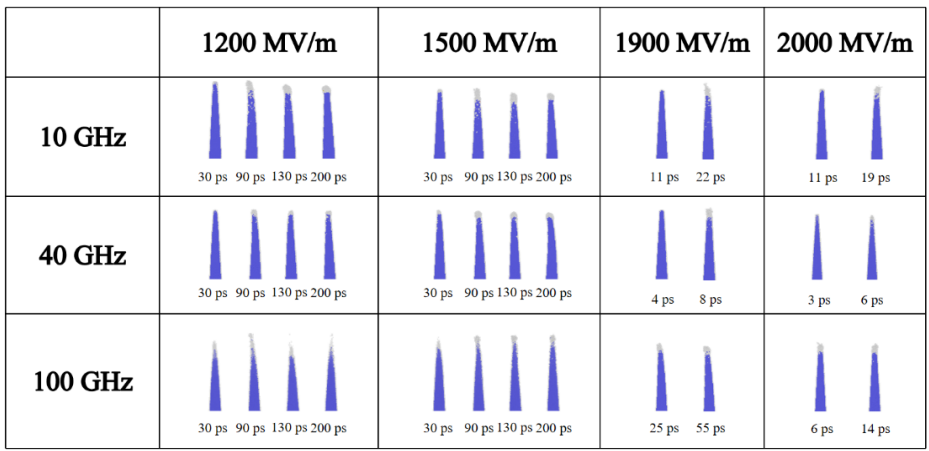

(c)

**Figure. 5** Snapshots representing atomic structure evolutions of Cu, Ti and W nanotips with a radius of curvature of 1 nm under the radiofrequency electric fields with various frequencies and field amplitudes. Colors illustrate the local coordination numbers of atoms, and red lines highlight the stacking faults. Note that the axial direction of Cu, Ti and W nanotips are oriented in [001], [0001] and [110] directions, respectively.

The atomic structure evolutions of Cu, Ti and W nanotips with the radius of curvature of 5 nm are displayed in Figure. 6 under the radiofrequency electric fields. The large metal nanotips deform drastically differently to those of small nanotips in terms of the changing of geometry and the extent of phase transformation with the heat and electric field. For example, the bending or tilting behavior is not seen in Figure. 6 for Cu, Ti and W nanotips with $r_0$ = 5 nm, compared to structural deformations presented in Figure. 5 for the tips with a small radius of curvature ($r_0$ = 1nm). Additionally, the apex of the large tips also experiences the phase transformation from crystalline solid to liquid phase, but the molten region is not substantially elongated with the applied electric field especially in the cases of Ti and W tips. When the applied electric field frequency is set to 10 GHz, both Cu and Ti nanotips show substantial melting of apex region by increasing the electric filed amplitude. Otherwise, Cu nanotip could be subjected to stronger deformation in comparison to that of Ti and W in the molten region.

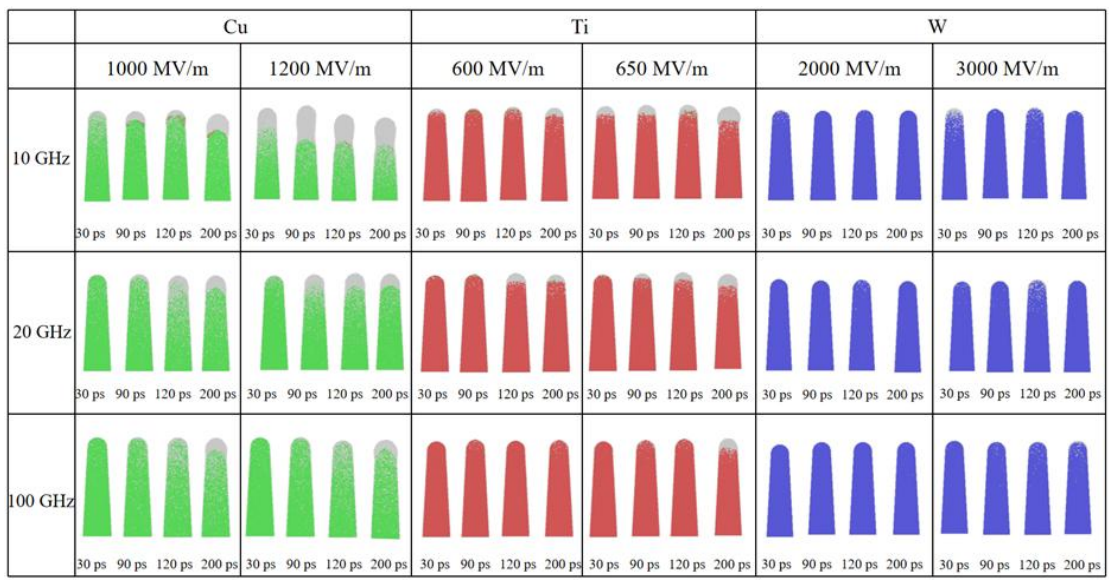

**Figure.6** Representing snapshots showing the structural evolutions of Cu, Ti and W nanotips with a radius of curvature of 5 nm under the radiofrequency electric fields.

Besides the observed apparent structure deformation and phase transformation processes, various point defects, dislocation lines, and stacking faults are created in the interior of large metal nanotips. Some characteristic structural defects of Cu, Ti and W nanotips with a radius of curvature of 5 nm under the radiofrequency are highlighted in Figure. 7. In the case of Cu nanotip, it is revealed that a large number of atomic defects are generated during the heating and cooling cycle with the applied radiofrequency electric field, including local atomic registry disordering, the stacking faults at {111} planes and dislocations of types 1/6 <112> (Shockley partial dislocation at {111} plane) and 1/6 <110> (Lomer-Cottrell dislocation at {111} plane). The point defects such as atomic registry disordering are widely distributed in the atomistic domain of Cu nanotip, while dislocations and stacking faults are largely confined in the upper section of nanotip, neighboring the molten apex. It is also seen that both dislocations and stacking faults coexist in the same volume in Cu nanotip, indicating their formation is closely related to each other, as shown in Figure. 7(a). Similar to those of Cu nanotip, the types of atomic defects and their locations are displayed in Figure. 7(b) for Ti nanotip with a radius of curvature of 5 nm. Clearly, the number of point defects is substantially less than that of Cu nanotip. Otherwise, the stacking faults at (0001) basal plane of Ti nanotip are oriented horizontally in the radial direction. By the meantime, the Schockley partial dislocations (1/3 [1-100]) in the basal (0001) plane are also aligned in the horizontal direction. Both stacking faults and dislocation lines are not seen in the case of W nanotips, as shown in Figure. 7(c) with the applied electric fields. In addition, very few point defects are formed in the heating and cooling cycle under the radiofrequency electric fields in the interior of the W nanotip with a radius of curvature of 5 nm, compared to that of either Cu or Ti case. Overall, we find that for refractory metal with a high melting point and strong interatomic cohesion forces, the formation of various point and line defects is greatly reduced, and structural deformation of nanotip under electric field shows high rigidity. In addition, thermal runaway proceeds through the formation of atomistic filaments at sub-nanoscale on the surface of metal nanotip for refractory metals such as W studied in current paper.

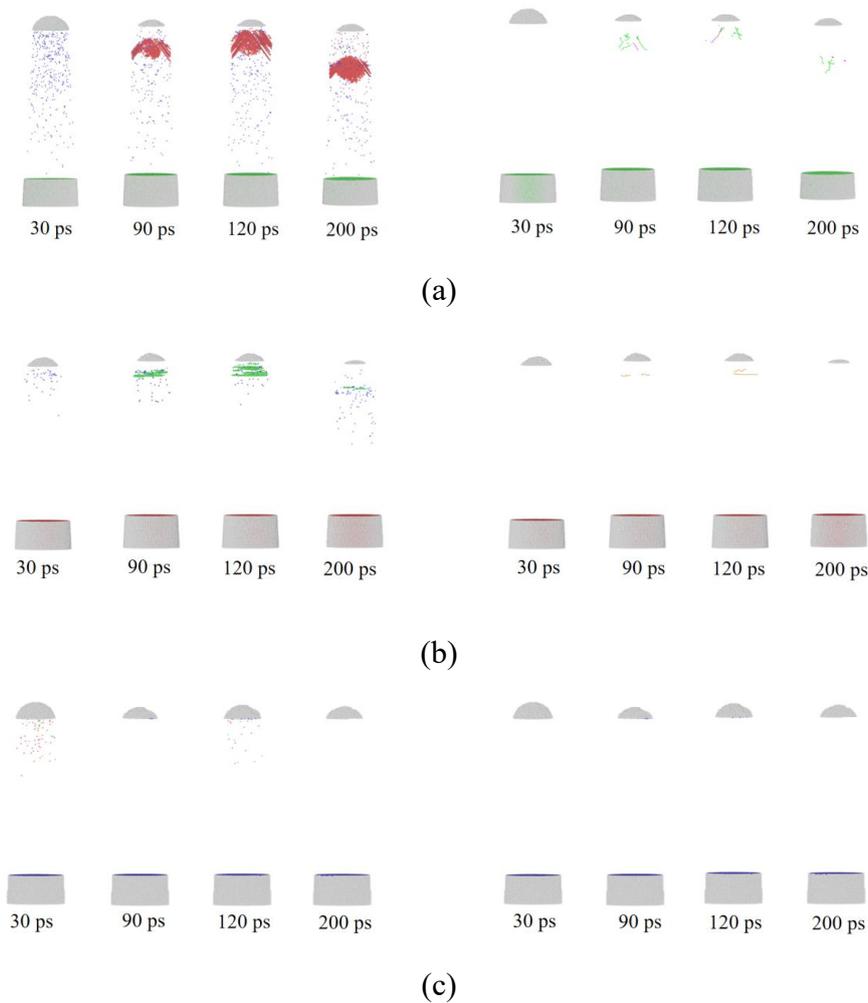

**Figure. 7** The structure defects of metal nanotips ($r_0$ = 5 nm) in ED-MD simulations under radiofrequency electric fields: (a): Cu, 1000 MV/m and 10 GHz; (b): Ti, 650 MV/m and 10 GHz; (c): W, 3000 MV/m and 10 GHz. The dots represent the atomic local registry disordering, the colored atomic layers indicate the stacking faults at {111} and (0001) crystallographic planes for FCC-Cu and HCP-Ti, respectively; the wavy lines are referred to dislocations.

## 3.2 Critical timescale of thermal runaway in ED-MD simulations

The typical timescales for observing the thermal runaway in ED-MD simulations for Cu, Ti and W metal nanotips with a radius of curvature of 1 nm under the applied radiofrequency electric fields are plotted in Figure. 8. The similar results are not shown for the nanotips with a larger radius at the apex ($r_0$ = 5 nm), because thermal runaway event is not predicted for most atomistic models within a maximum MD duration of 200 ps in current work. For all plots, we clearly see a strong non-monotonical dependence of critical timescale with the frequency ($f$) of the electric field for the thermal runaway, and the presence of a flat basin in those three-dimensional contour plots highlights the existence of the most feasible heating-cooling conditions to give the shortest time delay for thermal runaway event in the atomistic domain. In the case of Cu nanotip, as shown in Figure. 8(a), the critical E-field amplitude is somewhere between 500 MV/m and 650 MV/m for observing the thermal runaway process, regardless of the electric field frequencies (10 GHz, 40 GHz and 100 GHz). Meanwhile, the shortest time delay (10 ~ 20 ps) is obtained at around 40 GHz when the applied electric field amplitude is about 650 MV/m or above. Otherwise, the smallest critical timescale (~ 15 ps) of Ti nanotip exhibits a wide flat minimum basin for the frequency in a range from 25 GHz to 50 GHz when the applied electric field amplitude is above 250 MV/m, as shown in Figure. 8(b). The shortest time delay (~ 10 ps) of thermal runaway for W nanotip with a radius of curvature of 1 nm is found to be somewhere between 45 GHz and 55 GHz. It is worth noting that the observed dependence of critical time scale in thermal runaway for Cu, Ti and W nanotips is mainly determined by the accumulation of heat in the phonon subsystem in the two-temperature heat conduction model within the current ED-MD simulation methodology. Thus, the shortest time delay for the thermal runaway event in each metal nanotip is expected to change when the electron-phonon energy change rate ($G_{ep}$) is set to a different value. However, the non-linear correlation between critical time scale and the electric field frequency, as shown in Figure. 8 for all three metal nanotip with a radius curvature of 1 nm, remains valid with a shifting of the flat basin in frequency domain.

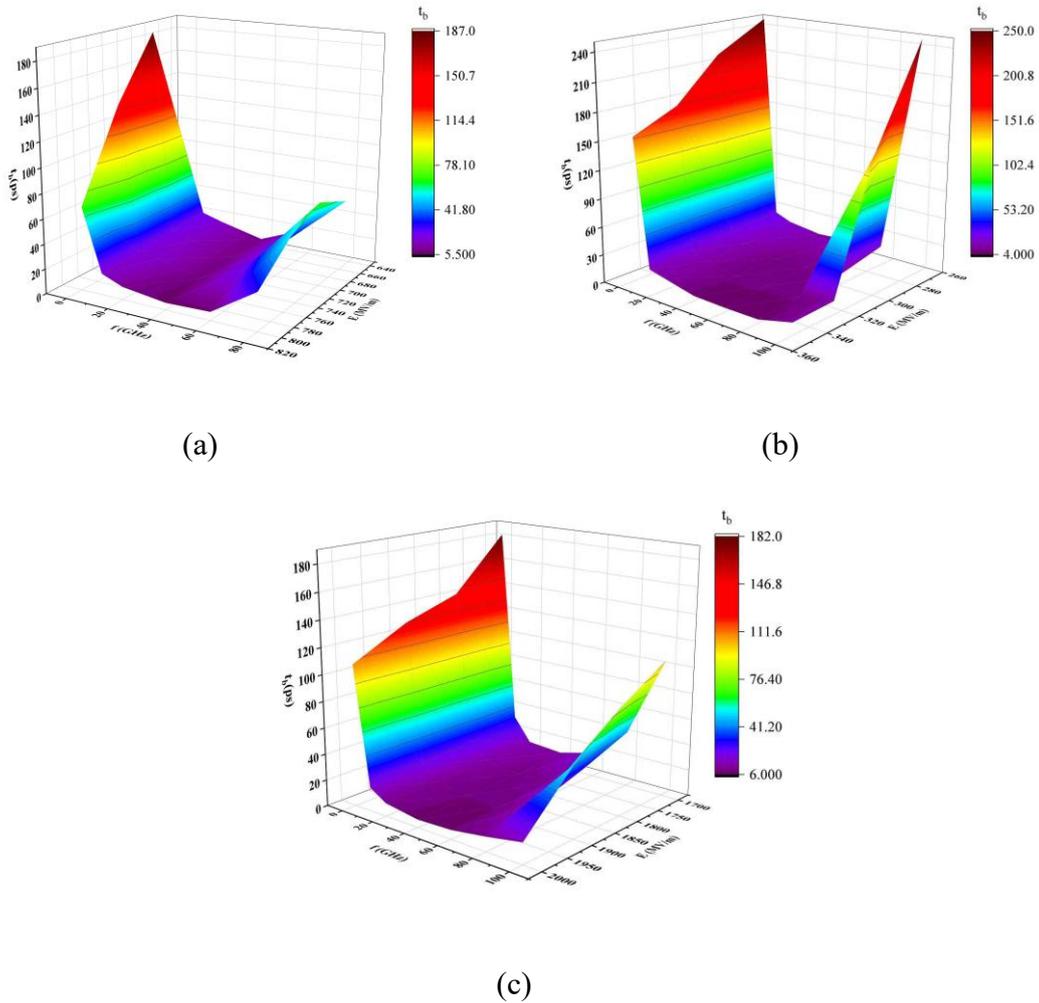

(a)    (b)

(c)

**Figure. 8** The obtained critical timescales for thermal runaway events of metal nanotips with a radius of curvature of 1 nm from ED-MD simulations for atomistic models under radiofrequency electric fields: (a): FCC-Cu; (b): HCP-Ti; (c): BCC-W.

By increasing the radius of curvature of Cu, Ti and W nanotips from 1 nm to 5 nm, the drastic change of critical time scale and electric field amplitude is obtained. First of all, we find that observing the melt at the apex in large Cu and Ti nanotips ($r_0$ = 5 nm) requires much higher electric field amplitudes than those of small nanotips ($r_0$ = 1 nm), i.e., 400 MV/m versus 1000 MV/m for Cu and 200 MV/m versus 600 MV/m for Ti. Otherwise, the large W nanotips ($r_0$ = 5 nm) show very little structural changes with the applied electric fields of 2 GV/m and 3 GV/m even at the apex regardless of the field frequencies, implying the relatively weak thermal effects due to Joule heating mechanism. In general, the large metal nanotips with a radius of curvature as 5 nm could sustain much higher external electric field in radiofrequency range, compared to the small nanotips ($r_0$ = 1 nm). The main reason is that the local electric field enhancement factor of large nanotip is reduced according to Eq. (4) for a paraboloid shape, i.e., $\beta$ = 144.7 ($d$ = 1 μm, $r_0$ = 1 nm) and $\beta$ = 37.7 ($d$ = 1 μm, $r_0$ = 5 nm). As a result, nanotip with a large radius of curvature exhibits a blunt apex, weakening the field electron emission process and the resulting heating mechanisms. For metal nanotips presented in Figure. 4, thermal runaway event is not detected for most cases within the maximum ED-MD simulation length as 200 ps. The only exception is seen in Figure. 4 for Ti nanotip with an electric field amplitude of 650 MV/m and the frequency as 100 GHz, the thermal runaway indeed occurs after 120 ps in ED-MD simulation. In summary, the critical time scale of thermal runaway in ED-MD simulations for Cu, Ti and W nanotips with a radius of curvature of 5 nm is revealed to be higher than 200 ps in most situations under the applied radiofrequency electric fields.

$$\beta = \frac{\dfrac{d}{r}}{\ln\left(\dfrac{d}{r}\right)} \qquad (4)$$

## 3.3 Instantaneous mass density profiles of nanotips

Here, we first illustrate the calculated mass density profiles ($z$-profiles) in the axial direction of metal nanotips that are subjected to the severe melting and crystallization processes during the heating-cooling cycles under the applied radiofrequency electric field. For computing mass density profiles, Eqs. (1)-(3) are employed for all snapshots in an ED-MD trajectory using previously proposed coarse-graining scheme. The results for six representative cases of Cu, Ti and W nanotips are displayed in Figure. 9. The extraction of the melt in each snapshot from ED-MD simulation is realized using an in-house Python automation script interfaced with OVITO software, based on the workflow shown in Figure. 4. In Figures. 9(a), 9(b) and 9(c), the evolutions of the nano-melts in the axial direction of Cu, Ti and W nanotips with a radius of curvature of 1 nm are demonstrated together with the calculated mean mass density and temperature of the molten region. Meanwhile, the similar results for the metal nanotips with a radius of curvature of 5 nm are depicted in Figures. 9(d), 9(e) and 9(f) for Cu, Ti and W, respectively. It is worth noting that all results presented in Figure. 9 for metal nanotips are subjected to different electric filed amplitudes and frequencies, thus the corresponding melting and recrystallization cycle are expected to be also different. Otherwise, for all extracted melts, their representative atomistic models are given in Figure. 10 at different timesteps under the same electric field conditions as those of Figure. 9 for Cu, Ti and W nanotips.

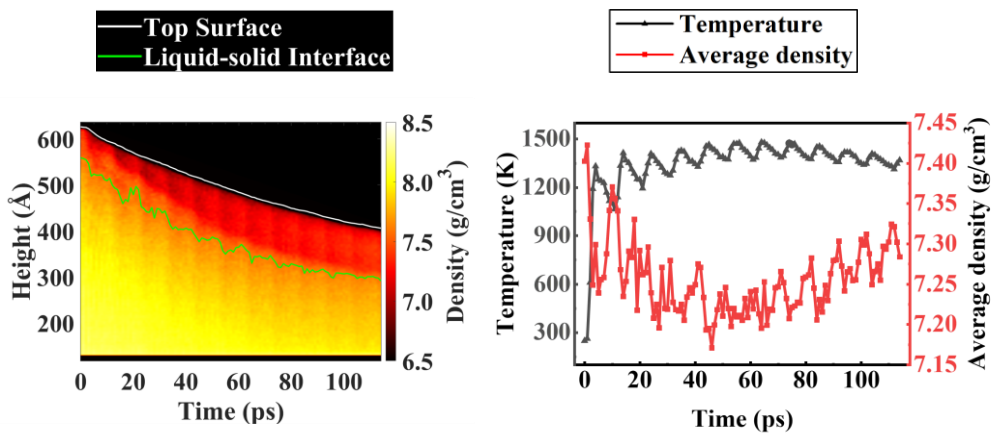

(a)

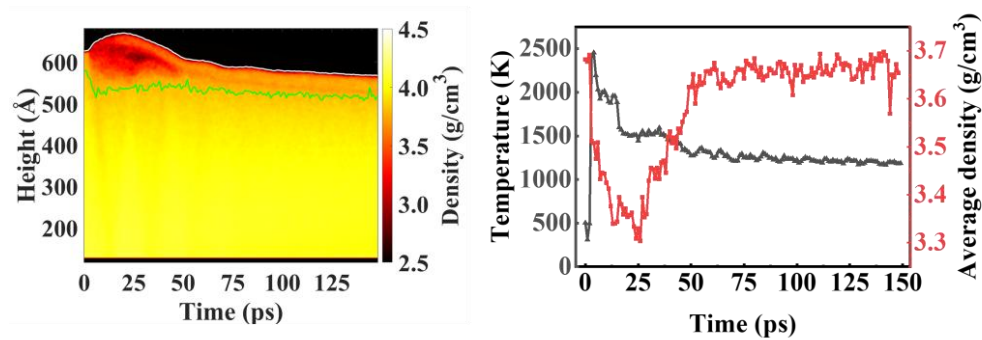

(b)

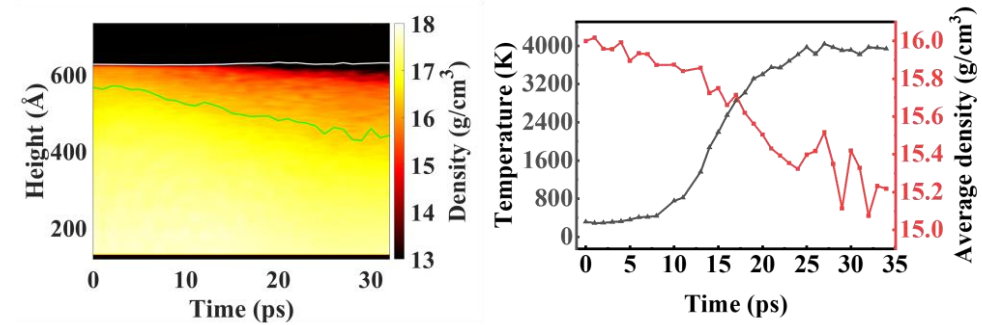

(c)

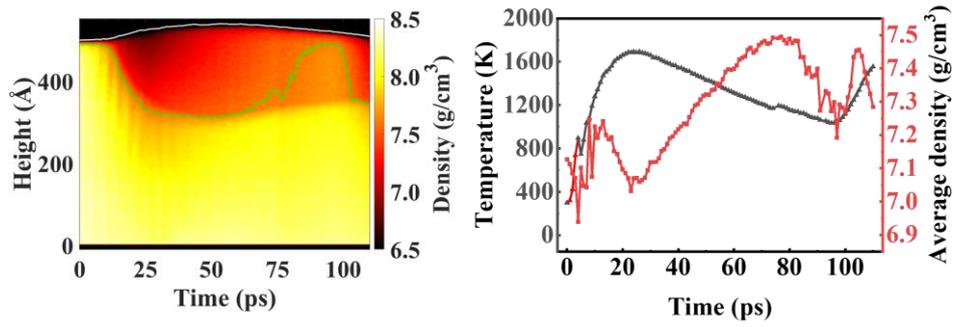

(d)

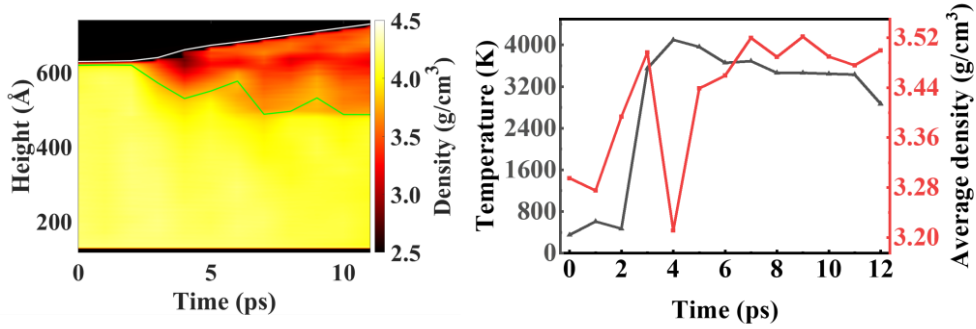

(e)

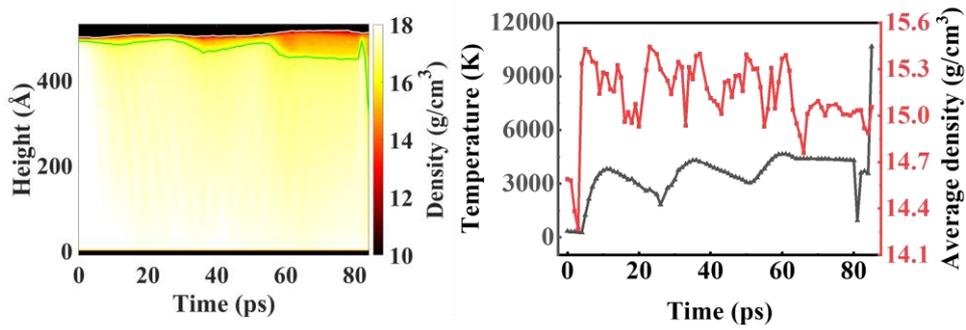

(f)

**Figure. 9** The instantaneous mass density profiles in the axial direction of metal nanotips and the associated mean mass density and temperature of the melt under radiofrequency electric fields: (a): Cu, $r_0$ = 1 nm, $E$ = 800 MV/m and $f$ = 100 GHz; (b): Ti, $r_0$ = 1 nm, $E$ = 300 MV/m and $f$ = 100 GHz; (c): W, $r_0$ = 1 nm, $E$ = 1900 MV/m and $f$ = 10 GHz; (d): Cu, $r_0$ = 5 nm, $E$ = 1200 MV/m and $f$ = 10 GHz; (e): Ti, $r_0$ = 5 nm, $E$ = 1200 MV/m and $f$ = 100 GHz; (f): W, $r_0$ = 5 nm, $E$ = 6000 MV/m and $f$ = 40 GHz. In all contour plots, the dark region denotes the vacuum or the vapor.

For Cu nanotips, we predict a drastic change in the tip geometry from the initial conic shape to a severe blunted mushroom-head-like apex for the smaller tips with the radius of curvature of 1 nm, as shown in Figure. 10(a) for the atomistic models under various radiofrequency electric fields. A significant reduction of the height of Cu nanotip is seen in Figure. 10(a), i.e., ~ 20 nm, which fully agrees with the decreasing of the tip height in the superimposed instantaneous mass density profiles illustrated in Figure. 9(a). The mushroom-head retains a molten state during multiple consecutive heating and cooling cycles with an electric field frequency as 100 GHz and the amplitude of 800 MV/m for a total duration of 114 ps in ED-MD simulation, as also evidenced from the transient temperature profile presented in the same graph. In the case of Cu nanotip with a radius of curvature of 5 nm, the expansion of melt at the apex is easily seen in Figure. 9(d), and the associated evolution of the atomistic model is referred to Figure. 10(d) with $f$ = 10 GHz and $E$ = 1200 MV/m. The atomistic model clearly indicates the molten region undergoes a necking process with the applied electric stress, favoring the formation of blunted apex with a semi-spherical or mushroom-head like geometry during the cooling stage. Obviously, reducing the sharpness of the metal nanotip could reduce the local electric field strength and field electron emission process, compared to that of conic tip apex, eventually stabilizing the tip structure. The density of the melts for Cu nanotips in the cases of Figures. 9(a) and 9(d) is estimated as the mean value from all transient mass density profiles for the molten region, and the values are found to be 7.15 g/cm$^3$ and 7.38 g/cm$^3$, respectively, compared to 8.96 g/cm$^3$ of crystalline Cu metal and 7.90 g/cm$^3$ at the melting point (1084 ℃).[61] Notably, the right panel in Figure. 9(a) indicates that the temperature of the melt at the apex of Cu nanotip could reach around 1400 K ~ 1500 K for a relatively long time (~ 90 ps). At such a temperature, liquid Cu has a mass density of 7.75 g/cm$^3$,[61] and which is still higher than the obtained values of nano-melt under electric field. Therefore, the observed discrepancies between nano-melt and the bulk Cu liquid in mass density reflects the influence of local electric stress on the molten state at nanoscale.

In the case of smaller Ti nanotip ($r_0$ = 1 nm), the superimposed instantaneous mass density profiles elucidate an interesting atomic structure evolution scenario that the melt is formed shortly after applying the electric field, then its volume further expands, and later on, which experiences a significant elongation by the electric stress (See Figure. 9(b)). However, the stretched molten region is contracted after the disappearance of the electric stress in the negative half period, and the melt eventually is merged with the bottom crystalline region in the cooling stage, as shown in Figure. 10(b). For the larger Ti nanotip ($r_0$ = 5 nm), we see the melt is stretched upward continuously during the whole ED-MD simulation from Figure. 9(e), while a substantial expansion of the volume of the melt is also revealed at the apex. Furthermore, the elongation of the molten region is proceeded through the necking of the melt and the blunting of the apex with the electric stress, as can be seen from Figure. 10(e). The mean densities of the melts for Ti nanotips are calculated to be 3.31 g/cm$^3$ and 3.56 g/cm$^3$ for $r_0$ = 1 nm and $r_0$ = 5 nm, respectively. Therefore, mass density of nano-melts under the electric stress is significantly less than that of either crystalline Ti (4.54 g/cm$^3$) or the liquid state at melting point (1941 K, 4.21 g/cm$^3$).[62,63] Extrapolating the density of liquid Ti above 2000 K using the experimental formula in Ref. 62 gives 4.01 g/cm$^3$ (2500 K) and 3.68 g/cm$^3$ (3500 K), respectively. Interestingly, the transient mass density of Ti nano-melt is revealed to be around 3.50 g/cm$^3$ when the instantaneous temperature reaches 3500

K, as can be seen from the right panel in Figure. 9(e) for $r_0$ = 5 nm. This value of nano-melt is closer to the extrapolated mass density of liquid Ti metal at the same temperature, though the nano-melt of Ti remains less dense than that of bulk liquid. Otherwise, great cautious should be paid to the extrapolated liquid density of Ti at very high temperatures, because the experimental measurements were only conducted in a very narrow temperature window (< 2100 K) above the melting point (1941 K) within the electromagnetic levitation apparatus in the literature.[62,63]

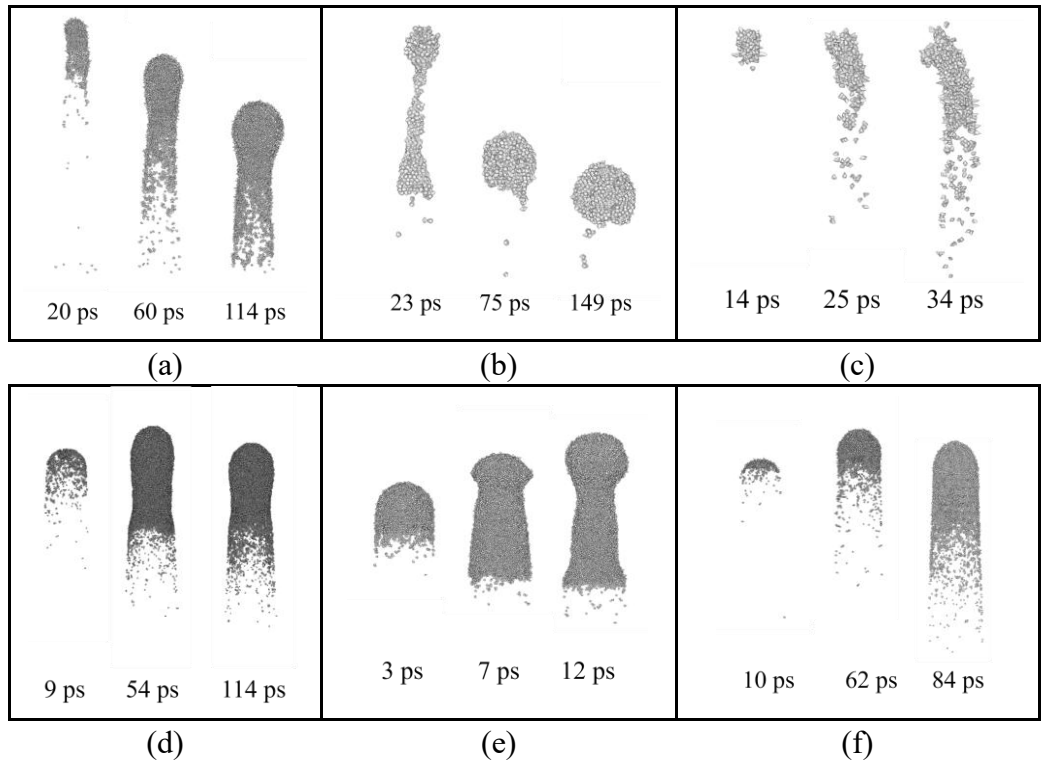

**Figure. 10** Representative atomistic models of the extracted melts from trajectories of ED-MD simulations for metal nanotips at different timesteps: (a): Cu, $r_0$ = 1 nm, $E$ = 800 MV/m and $f$ = 100 GHz; (b): Ti, $r_0$ = 1 nm, $E$ = 300 MV/m and $f$ = 100 GHz; (c): W, $r_0$ = 1 nm, $E$ = 1900 MV/m and $f$ = 10 GHz; (d): Cu, $r_0$ = 5 nm, $E$ = 1200 MV/m and $f$ = 10 GHz; (e): Ti, $r_0$ = 5 nm, $E$ = 1200 MV/m and $f$ = 100 GHz; (f): W, $r_0$ = 5 nm, $E$ = 6000 MV/m and $f$ = 40 GHz.

For W nanotips, regardless of their radii of curvature, substantial elongation of the molten region at the apex is seen neither in Figure. 9(c) ($r_0$ = 1 nm) nor Figure. 9(f) ($r_0$ = 5 nm). As a result, the height of nanotip shows very little variation during the ED-MD simulations. However, further inspecting the mass density profiles of both tips indeed indicates the increase of the volume of molten region due to the heating processes with the applied electric fields. The atomistic models of the extracted molten regions at different timesteps are displayed in Figure. 10(c) and 10(f) for $r_0 = $ 1nm and $r_0 = $ 5 nm, respectively. The results clearly show the expansion of the melt at the apex towards the bottom of the nanotip in the ED-MD simulation. Additionally, for smaller W nanotip ($r_0 = $ 1 nm), the high electric stress causes the further sharpening of the molten region in terms of the formation of spiky morphology at atomic scale shortly before the occurring of thermal runaway event. The transient temperature in the molten region can reach 3000 K ~ 4000 K for W nanotips, as shown in the right panels of Figures. 9(c) and 9(f). The mean densities of the melts for W nanotips are estimated to be 14.81 g/cm$^3$ ($r_0 = $ 1nm) and 14.90 g/cm$^3$ ($r_0 = $ 5 nm), respectively. For a comparison, the densities of crystalline W at room temperature and liquid state at the melting point (3695 K) are reported as 19.25 g/cm$^3$ and 16.267 g/cm$^3$, respectively.[64] The corrected mass density of liquid W at 4000 K is about 16.02 g/cm$^3$, remaining considerably larger than the densities of nano-melts at the apex of nanotips.[64,65]

In general, we find all melts at the apex of nanotips exhibit substantially smaller mass densities than those of either crystalline phases for Cu, Ti and W metals or their corresponding bulk liquids at high temperature. However, the obtained mass densities of the molten apexes in those metal nanotips are also well above the critical values in the liquid-vapor coexisting state of Cu, Ti and W elements, i.e., 1.895 g/cm$^3$ ($T_c = $ 6550 K),[62,66] 2.60 g/cm$^3$ ($T_c \sim $ 6600 K),[67] and 4.2 g/cm$^3$ ($T_c = $ 12200 K).[65,68] Thus, the molten apex is considered as the nano-melt that is subjected to high electric stress and thermal dilation locally, substantially decreasing its mass density, compared to that of conventional liquid metal.

## 3.4 Evolution of kinematic viscosities of nanomelts at tip apex

For calculating both dynamic and kinematic viscosities of the melts for Cu, Ti and W nanotips, all atoms in the molten region at the apex are extracted for each snapshot from the ED-MD trajectory using the workflow presented in Figure. 4. Then, Eq. (4) is employed to evaluate transient dynamics viscosities of melts with the rf electric fields from the atomistic simulations. When applying Eq. (5) to convert dynamic viscosity into the kinematic quantity, the transient mass density profile is adopted, as displayed in right panel of Figure. 9 for each metal nanotip and rf electric field conditions. Since the obtained dynamic viscosity profiles closely resemble those of kinematic results, only the latter quantity is addressed in the current paper in each case. Typical results are shown in Figure. 11 for the relevant metal nanotips under various rf electric field conditions. For all nano-melts, the calculated kinematic viscosities in axial direction (or parallel to the applied electric field) are significantly larger than those in the radial direction (or perpendicular to the applied electric field), i.e., $v_{//} \approx 10\sim100\ v_{\perp}$. This observation is largely attributed to the long-range particle migration in the molten region toward the axial direction during the severe structural deformation processes such as necking and elongation of melts and blunting of tip apex under the rf electric field stress especially when thermal runaway is seen in the ED-MD simulation for atomistic models. Those types of structural deformation require both mass and momentum transportations in the direction of electric stress. Otherwise, atoms are expected to be confined in radial direction as long as the structural integrity is retained for metal nanotips in ED-MD simulations. As a result, the kinematic viscosity in the radial direction may be less affected by the electric field stress, and its value should be close to that of bulk liquid metal.

The variations of kinematic viscosities of nano-melts for Cu nanotips with the radii of curvature of 1 nm and 5 nm are plotted in Figures. 11(a) and 11(d), respectively. In the case of Figure. 11(a), the strong fluctuations in the obtained kinematic viscosity versus time profiles are related to the multiple consecutive heating-cooling cycles during the whole ED-MD simulations with a rf value of 100 GHz. For both Cu nanotips, the mean kinematic viscosities are well-defined for $v_{//}$ and $v_{\perp}$ components, because all transient profiles exhibit the flat region with the time. For small nanotip with the radius of curvature of 1 nm, the mean values of $v_{//}$ and $v_{\perp}$ components are found to be $(8.05 \pm 1.01) \times 10^{-5}$ m$^2$/s and $(7.56 \pm 1.01) \times 10^{-6}$ m$^2$/s, respectively. Meanwhile, the values for the melt of large Cu nanotip ($r_0 = 5$ nm) were obtained as $v_{//} = (5.49 \pm 0.09) \times 10^{-4}$ m$^2$/s and $v_{\perp} = (3.86 \pm 0.15) \times 10^{-4}$ m$^2$/s. The kinematic viscosity of bulk liquid Cu reported in Refs. 61,69,70 for the temperature ranging from the melting point (~ 1357 K) to 1673 K are given as $5.06 \times 10^{-7}$ m$^2$/s ~ $3.20 \times 10^{-7}$ m$^2$/s. It is interesting to see that the calculated kinematic viscosities ($v_{//}$ and $v_{\perp}$) of melts for small Cu nanotip more resemble that of bulk Cu liquid, compared to those numbers obtained for the tip with a large radius of curvature. Nevertheless, for both Cu nanotips, the nano-melts at the apex are expected to behave differently to the conventional liquid Cu metal without applying rf electric field, as the obtained kinematic viscosities are at least one order of magnitude higher than that of ordinary case.

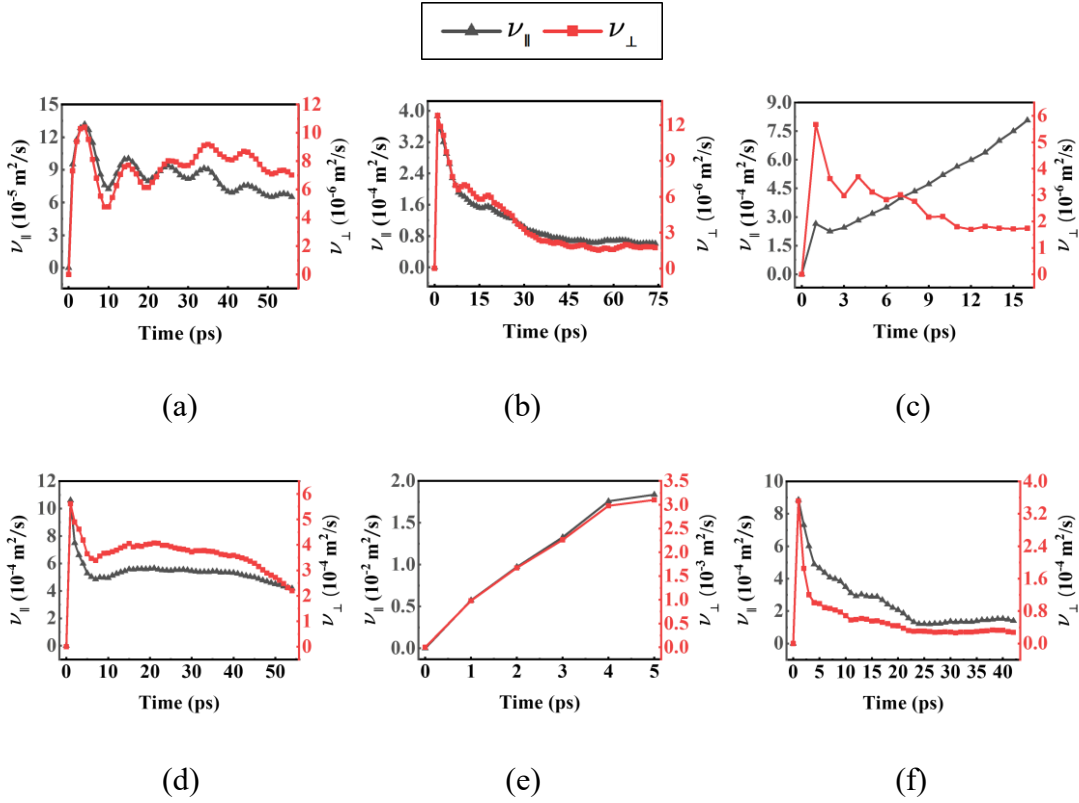

**Figure. 11** The calculated transient kinematic viscosities of nanomelts at the apex of metal nanotips with the applied rf electric fields: (a): Cu, $r_0 = 1$ nm, $E = 800$ MV/m and $f = 100$ GHz; (b): Ti, $r_0 = 1$ nm, $E = 300$ MV/m and $f = 100$ GHz; (c): W, $r_0 = 1$ nm, $E = 1900$ MV/m and $f = 10$ GHz; (d): Cu, $r_0 = 5$ nm, $E = 1200$ MV/m and $f = 10$ GHz; (e): Ti, $r_0 = 5$ nm, $E = 1200$ MV/m and $f = 100$ GHz; (f): W, $r_0 = 5$ nm, $E = 6000$ MV/m and $f = 40$ GHz. The viscosity tensor components are defined as $v_{//} = (v_{xz} + v_{yz})/2$ and $v_\perp = v_{xy}$ in the axial and radial directions.

The results for the two representative cases of Ti nanotips are plotted in Figures. 11(b) and 11(c) with the radii of curvature of 1 nm and 5 nm, respectively. Notably, the applied rf frequency is the same (100 GHz) in those two examples, though the applied electric field amplitudes are different, i.e., 300 MV/m ($r_0 = 1$ nm) and 1200 MV/m ($r_0 = 5$ nm). For the small Ti nanotip, both $v_{//}$ and $v_\perp$ profiles are flattened after 35 ps, and the reason is that thermal runaway is not predicted for the nanotip under such rf electric field conditions ($f = 100$ GHz and $E_{max} = 300$ MV/m) and the atomic model is stabilized especially after the blunting of the apex through melting and crystallization processes. Meanwhile, the plots for the large tip ($r_0$ = 5 nm) increase monotonically with the whole simulation time, and in fact experiences the thermal runaway after 12 ps in ED-MD simulation with the rf electric field amplitude 1200 MV/m. Therefore, the melt of the apex for the large Ti nanotip is subjected to vary large local electric stress and which is relatively short-lived. Thus, for computing the mean kinematic viscosity of large Ti nanotip, the average is taken for the whole profile of either $v_{//}$ or $v_\perp$ component. Since the ED-MD simulation time is rather short in this particular case, the obtained kinematic viscosities ($v_{//}$ and $v_\perp$) should be regarded as some rough estimations of the property. For small Ti nanotip, the obtained values of $v_{//}$ and $v_\perp$ components are $(6.86 \pm 0.67) \times 10^{-5}$ m$^2$/s and $(1.88 \pm 0.22) \times 10^{-6}$ m$^2$/s, while those for large tip are given as $(1.29 \pm 0.53) \times 10^{-2}$ m$^2$/s and $(2.20 \pm 0.89) \times 10^{-3}$ m$^2$/s, respectively. For a comparison, experimental values of liquid Ti for the temperature between the melting point (1943 K) and 2200 K are referred to $1.51 \times 10^{-7}$ m$^2$/s ~ $1.70 \times 10^{-7}$ m$^2$/s in Ref. 71, while Ref. 72 reports a different value as $1.06 \times 10^{-6}$ m$^2$/s at the melting point of Ti.

Finally, the variations of $v_{//}$ and $v_\perp$ components with the ED-MD simulation time for the two representative cases of W nanotips are illustrated in Figures. 11(c) and 11(f). For the small nanotip ($r_0$ = 1 nm), thermal runaway is observed at about 22 ps in the simulation with the rf frequency 10 GHz and the electric field amplitude 1900 MV/m. As a result, we see the monotonic change of the $v_{//}$ profile versus the time. Otherwise, the radial component $v_\perp$ actually shows a flat region after 12 ps, and which is notably different to the profiles shown in Figure. 11(e) for Ti nanotip ($r_0$ = 5 nm). For large W nanotip, both $v_\parallel$ and $v_\perp$ components are well converged after 20 ps in the ED-MD simulation under the applied rf electric field conditions (rf frequency of 40 GHz and the electric field amplitude 6000 MV/m). The mean kinematic viscosities of the melt for small W nanotip are calculated to be $(4.73 \pm 1.88) \times 10^{-4}$ m$^2$/s and $(1.75 \pm 0.04) \times 10^{-6}$ m$^2$/s, and the numbers of large tip are given by $(1.35 \pm 0.11) \times 10^{-4}$ m$^2$/s and $(2.95 \pm 0.19) \times 10^{-5}$ m$^2$/s. Regarding the experimental values, the kinematic viscosity of liquid W is measured to be $5.21 \times 10^{-7}$ m$^2$/s at the melting point (3695 K), and the extrapolated value at 6000 K is about $1.31 \times 10^{-7}$ m$^2$/s.[64] Again, melts of W nanotips exhibit higher kinematic viscosity than that of ordinary W liquid.

We summarize the calculated kinematic viscosities ($v_\parallel$ and $v_\perp$) for the six representative ED-MD simulations of Cu, Ti and W metal nanotips in Table. 2 together with the mean values of mass density and temperature for each case. Overall, the kinematic viscosities of the melt at the apex of metal nanotips are substantially higher than that of experimental values for the bulk liquid metal by orders of magnitude especially for the viscosity tensor component along the axial direction where the melt is expected to experience the strong electric stress. In some cases, due to the thermal runaway in the ED-MD simulations, the obtained kinematic viscosities are enormously large such as Cu ($r_0 = 5$ nm), Ti ($r_0 = 5$ nm) and W ($r_0 = 1$ nm), compared to the corresponding experimental values for liquid metal. The calculated mean temperature for each case is either slightly below the melting point of bulk metal or approximate to the melting point, implying the possibility that the melt could be transformed into the under-cooled liquid phase or simply the solidified amorphous state during the multiple heating-cooling cycles with rf electric fields. Nevertheless, the drastic differences in the values of kinematic viscosity between radial and axial components reflect the strong influence of electric field stress on the fluid dynamics of nano-melts.

**Table. 2** The calculated kinematic viscosities of six representative nanotips and the comparison with experimental values at the melting point for each metal. The units for the relevant quantities are K, g/cm³ and m²/s for average temperature ($T_{ave}$), mass density ($\rho_{ave}$) and kinematic viscosity ($v$). Note that $v_{//} = (v_{xz} + v_{yz})/2$ and $v_{\perp} = v_{xy}$, referring to the tensor components in the axial and radial directions, respectively.

| Nanotip | $T_{ave}$ | $\rho_{ave}$ | $v_{//}$ (×10⁻⁶) | $v_{\perp}$ (×10⁻⁶) | $v_{exp}^{69,71,72,64}$ (×10⁻⁷) |
|---|---|---|---|---|---|
| Cu (1 nm) | 1342.9 | 7.15 | 80.48 ± 10.14 | 7.50 ± 1.01 | 5.06 |
| Cu (5 nm) | 1630.5 | 7.38 | 548.88 ± 9.20 | 385.58 ± 14.74 | |
| Ti (1 nm) | 1340.7 | 3.31 | 68.61 ± 6.72 | 1.88 ± 0.22 | 1.51 |
| Ti (5 nm) | 2173.8 | 3.56 | ~12930 | ~2200 | 10.60 |
| W (1 nm) | 1943.0 | 14.81 | 473.06 ± 188.57 | 1.75 ± 0.04 | 5.21 |
| W (5 nm) | 3603.2 | 14.90 | 135.64 ± 11.17 | 29.48 ± 1.93 | |

## 3.5 Critical spatial and temporal scales of instability for electrocapillary waves of nanomelts

The instability of electrocapillary waves with the consideration of kinematic viscosity of the liquid metal fluid in the dispersion equation exhibits two distinct regimes known as the ideal fluid limit and the viscosity-dominated regime. The boundary between those two regimes is determined by the local critical electric field value ($E_c$) that is estimated from the fundamental physical properties of the melt including the kinematic viscosity ($v$), surface tension ($\alpha$) and mass density ($\rho$), as well as the dielectric permittivity of vacuum ($\varepsilon_0$), as given by Eq. (5).[40] The average kinematic viscosity is used in all calculations, and the value is obtained from Eq. (6) from the axial and radial components.

$$E_c \approx \frac{\alpha}{\varepsilon_0^{1/2} \rho^{1/2} \nu} \quad (5)$$

$$\nu = \frac{2}{3}\nu_{//} + \frac{1}{3}\nu_{\perp} \quad (6)$$

Employing the relevant parameters of nano-melts for Cu, Ti and W nanotips from ED-MD simulations with various rf electric field conditions enable the estimation of the critical electric field strength for each case presented in Table. 2, and all results are summarized in Table. 3. As can be seen from Table. 3, the estimated values of the critical local electric field strength for various nano-melts under rf fields are usually in the range of $10^7$ M/m ~ $10^8$ V/m. An exception is the value of Ti nanotip with a radius of curvature of 5 nm, and critical electric field strength is found to be only 3 MV/m, and which is in fact several orders of magnitude less than either the applied rf field (1200 MV/m) or the transient local field value (23.8 GV/m). It is quite obvious that the very higher kinematic viscosity drastically lowers the critical electric field value in this particular case. Nevertheless, the applied rf electric fields and the local transient fields are well above the calculated critical electric field strength for each nanomelt listed in Table. 3, the viscosity-dominated regime is concluded for the observed instability of molten apex in ED-MD simulations.

We further evaluate the critical temporal and time scales for the dominate instability mode in the viscosity-dominated regime for nanomelt using Eqs. (7) and (8), respectively. Here, $k_d$ and $\gamma_d$ represent the dominate instability wave number and frequency, while their inversions give the critical wavelength ($\lambda_d$) and time scale ($\tau_d$).

The critical time scale may be directly compared to the predicted time delay of thermal runaway in ED-MD simulation for the nanotip.

**Table. 3** The calculated critical parameters of electrocapillary wave instability and other relevant quantities for Cu, Ti and W nanotips with rf electric fields, including rf electric field amplitude ($|E_0|$, MV/m), transient local maximum electric field strength ($E_{max}$, MV/m), surface tension ($\alpha$, N/m) at the melting point, density ($\rho_{ave}$, g/cm$^3$), kinematic viscosity ($\nu$, m$^2$/s), critical local electric field value ($E_c$, MV/m), critical electrocapillary wavelength in viscosity-dominated regime ($\lambda_d$, m) and the critical time scale for the dominate instability model ($\tau_d$, s). The parameter $\tau_0$ (ps) refers to the time delay to the thermal runaway from ED-MD simulations. Note that thermal runaway is not observed for Cu nanotip ($r_0$ = 5 nm) in the simulation with a maximum duration of 200 ps, the value is left empty with "—".

| Metal | $|E_0|$ | $E_{max}$ | $\alpha^{73,72}$ | $\rho$ | $\nu(\times 10^{-6})$ | $E_c$ | $\lambda_d$ | $\tau_d$ | $\tau_0$ |
|---|---|---|---|---|---|---|---|---|---|
| Cu (1 nm) | 800 | 10800 | | 7.15 | 18.72 | 285 | $1.0 \times 10^{-7}$ | $2.6 \times 10^{-10}$ | 114 |
| | | | 1.35 | | | | | | |
| Cu (5 nm) | 1200 | 181 | | 7.38 | 164.81 | 32 | $1.0 \times 10^{-4}$ | $8.4 \times 10^{-6}$ | — |
| Ti (1 nm) | 300 | 407 | | 3.31 | 15.45 | 589 | $5.7 \times 10^{-6}$ | $7.0 \times 10^{-8}$ | 149 |
| | | | 1.56 | | | | | | |
| Ti (5 nm) | 1200 | 23800 | | 3.56 | ~3117 | ~3 | $8.9 \times 10^{-7}$ | $4.4 \times 10^{-9}$ | 12 |
| W (1 nm) | 1900 | 96900 | | 14.81 | 315.96 | 21 | $5.6 \times 10^{-8}$ | $1.1 \times 10^{-10}$ | 22 |
| | | | 2.48 | | | | | | |
| W (5 nm) | 6000 | 75200 | | 14.90 | 100.25 | 68 | $3.6 \times 10^{-8}$ | $6.0 \times 10^{-11}$ | 79 |

$$\left. \begin{aligned} k_d &\approx \frac{3^{1/3} \varepsilon_0^{2/3} E^{4/3}}{2^{2/3} \nu^{2/3} \alpha^{1/3} \rho^{1/3}} \\ \lambda_d &= \frac{2\pi}{k_d} \end{aligned} \right\} \quad (7)$$

$$\left. \begin{aligned} \gamma_d &\approx \frac{\varepsilon_0 E^2}{2\nu\rho} \\ \tau_d &= \frac{1}{\gamma_d} \end{aligned} \right\} \quad (8)$$

In Table. 3, the obtained critical temporal and time scales for are presented for the six representative scenarios of Cu, Ti and W nanotips. By the meantime, the time delay of thermal runaway in ED-MD simulation is also listed for each nanotip. Regarding the critical wavelength of electrocapillary wave, the values are found to be between 10$^{-4}$ m

~ $10^{-8}$ m for the nanomelts with rf electric fields. Those results are orders of magnitude larger than the typical dimension of molten apex in a metal nanotip. Therefore, the development of instability is expected to happen for all metal nanotips at some time during the ED-MD simulations. The critical time scale for probably observing the instability of nanomelts is revealed to be several hundreds of picoseconds for W nanotips ($r_0$ = 1 nm and 5 nm), comparable to the time delays of thermal runway obtained from ED-MD simulations. Meanwhile, the critical time scales of Ti nanotips are typically longer than those of ED-MD simulations. In the case of Cu metal, the calculated critical time scale (~ 260 ps) of the small nanotip ($r_0$ = 1 nm) agrees with the time delay (114 ps) of thermal runaway in ED-MD simulation, while the instability time scale of the large nanotip ($r_0$ = 5 nm) is in the range of several microseconds. Notably, thermal runaway is not seen in ED-MD simulation for the large Cu nanotip within 200 ps duration.

Overall, both the critical spatial and time scales of molten region in W nanotips are fairly close to ED-MD results, if the thermal runaway of the melts are treated as the electrocapillary wave instability with the transient local electric field. Otherwise, substantial discrepancies are indeed seen in those quantities between electrocapillary instability and thermal runaway of ED-MD simulations in the cases of Cu and Ti nanotips with rf electric fields. In fact, as shown by Eqs. (7) and (8), two most important parameters that determine the critical spatial and temporal scales of instability of electrocapillary wave are the local electric field ($E$) and the kinematic viscosity ($v$) of nano-melt. Those two quantities could vary by many orders of magnitude for metal nano-protrusions under different electric field conditions. The high local electric field and low viscosity could drastically decrease the estimated critical wavelength and instability time scale for the melt under the electric field, i.e., Cu ($r_0$ =1 nm) and Ti ($r_0$ = 5 nm). In Ref. 40, the kinematic viscosity of Cu was on the order of $10^{-7}$ m$^2$/s, assuming the nanomelt behaves the same as the bulk liquid copper at the melting point. Employing such a low value of kinematic viscosity, the critical spatial and temporal scales of the instability for electrocapillary wave could be on orders of several nanometers and tens of picoseconds. In addition, the critical local electric field will be at least several GV/m for all metal nanotip in the viscosity-dominated regime. Those results appear to be fully consistent with the ED-MD simulations both for spatial and temporal scales of thermal runaway event. However, if the kinematic viscosity of nanomelt is indeed drastically different to that of ordinary liquid metal, then further investigation is needed to clarify deviations between our current estimates and those of Ref. 40.

## 4. Conclusion

In this paper, we have examined the electrohydrodynamic instability of metal nanotips consisting of Cu, Ti and W metals with different radii of curvature and under various rf electric field conditions within two different approaches, the electrodynamics coupled with molecular dynamics simulation (ED-MD) and the instability theory of electrocapillary wave with effects of local electric field and kinematic viscosity. The atomistic simulations of metal nanotips with radii of curvature of 1 nm and 5 nm under various rf electric field conditions indicate the melting and severe mechanical deformation of the nanomelts are strongly modulated by the field amplitude and frequency, while the thermal runaway is critically determined by the applied electric field strength that has distinctive value for each type of metal nanotip. Increasing the radius of curvature of nanotip generally results in higher critical electric field amplitude and longer time delay for observing thermal runaway in atomistic simulations. Current study also reveals the non-monotonic correlation between time to thermal runaway and the electric field frequency, and the existence of the minimum time delay for thermal runaway of the Cu, Ti and W nanotip with a radius of curvature of 1 nm at certain frequencies of rf fields implies a complex interplay among various factors including field electron emission induced heating processes, the hierarchical electron and phonon heat conduction mechanisms and the energy exchange rate between electron and phonon systems in ED-MD multiphysics-multiscale simulations using the atomistic models. Based on atomistic simulations, we have developed workflows to efficiently extract atoms of the molten apex in atomistic simulations from the ED-MD trajectory for computing the kinematic viscosity tensor components with Eistein-Helfand formula and the mass density of the nanomelts using a coarse-graining scheme. The obtained mass densities of nanomelts with the rf electric fields for Cu, Ti and W nanotips are revealed to be less than the bulk liquid metal, while the calculated kinematic viscosities of molten phase are usually orders of magnitude higher than those of conventional liquid metals at their melting points. Employing the instability theory for electrocapillary wave under the influence of both local electric field and liquid kinematic viscosity, critical electric field values are obtained for all metal nanotips under various rf electric field conditions, showing the viscosity-dominated instability mechanisms for all investigated cases. Notably, the estimated critical spatial and temporal scales of nanomelts within the viscosity-dominated instability regime for electrocapillary wave agree with the results of ED-MD simulations for W nanotips. However, drastic differences are found for the critical scales of the instability of nanomelts in the cases of Cu and Ti nanotips between electrocapillary wave dynamics and the ED-MD atomistic simulations, mainly because the kinematic viscosity obtained in the current work for all nanomelts with rf electric field are substantially higher than those of bulk liquid metals.

Our current work concludes that the influence of the applied electric field on the value of kinematic viscosity of nanomelt plays the most important role in determining the critical parameters for observing the instability of electrocapillary wave, requiring further thorough investigation in future study. Otherwise, workflows proposed and tested here bridge the methodological gap between the atomistic simulation and the continuum model under the principle of multiphysics-multiscale scheme for understanding the dynamics of nanomelt with the applied multiple physical fields.

**Acknowledgements**

This research is financially supported by the Young Talent Support Plan at Xi'an Jiaotong University awarded to Bing Xiao (No: DQ1J009) and the Fundamental Research Funds of the Central Universities (No. xtr052024009 and No. xyz022023092).